\documentclass[twocolumn,twoside]{svmultivs_gm} 
\usepackage{graphicx}
\title*{Current Status of Shanghai VLBI Correlator}
\titlerunning{Shanghai Correlator}
\author{Wu Jiang~$^{1,2}$, Zhiqiang Shen~$^1$, Fengchun Shu~$^1$, Zhong Chen~$^1$,Tianyu Jiang~$^1$}
\authorrunning{Jiang et al.} 
\authoremails{jiangwu@shao.ac.cn}
\institute{1. Shanghai Astronomical Observatory, Chinese Academy of Sciences \\
           2. University of Chinese Academy of Sciences }
\ContactAuthorName{Wu Jiang}
\ContactAuthorTelephone{+86-21-34775427}
\ContactAuthorEmail{jiangwu@shao.ac.cn}
\NumberofInstitutions{2}
\InstitutionPostAddress{1}{Shanghai 200030}
\InstitutionCountry{1}{China}
\InstitutionWebPage{1}{www.shao.ac.cn}
\InstitutionPostAddress{2}{ Beijing 100049}
\InstitutionCountry{2}{China}
\InstitutionWebPage{2}{N.A}
\begin{document}  
\maketitle       
\abstract{Shanghai Astronomical Observatory has upgraded its DiFX
cluster to a 420 cpu cores and a 432~TB storage system at the end of
2014. An international network connection for the raw data transfer
has also been established. The routine operations for IVS sessions
including CRF, AOV and APSG series began in early 2015. In addition 
to the IVS observations, the correlator is dedicated to astrophysical 
and astrometric programs with the Chinese VLBI network and international 
joint VLBI observations. It also worked with the new-built Tianma
65-m radio telescope and successfully found the fringes as high as at
X/Ka and Q bands in late 2015. A more powerful platform is planned for 
the high data rate and massive data correlation tasks in the future. }
%
%
%
\section{Introduction}
The VLBI group in Shanghai Astronomical Observatory (SHAO) has a long history of the development
with the VLBI correlator.The domestic software correlator and hardware correlator are mainly developed
and applied for the VLBI tracking system in the Chinese deep space missions. The worldwide open source software 
correlator called DiFX is adopt at SHAO in 2012 and works as a dedicated correlator for astrophysics 
and geodesy. The computer cluster and the data storage system of the DiFX correlator has been upgraded
in the end of 2014. It has 420 CPU cores and 432~TB storage capacity (Figure 1). An international high speed network
connection for the raw data transfer among main correlators and geodetic stations is established. 
Begining in 2015, the DiFX correlator is also served as an IVS correlator. By now, more than 10 IVS sessions such 
as CRF-, AOV-, APSG-, CRDS-series and a few Australian geodetic VLBI sessions have been processed by the plantform.
\begin{figure}[htb!]
  \includegraphics[width=.5\textwidth]{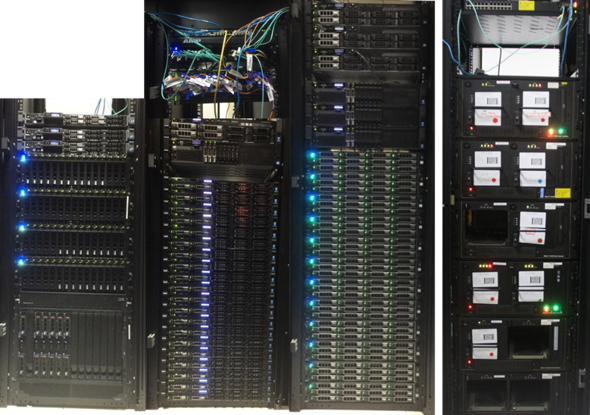}
  \caption{Hardware deployment of Shanghai VLBI correlator.}
  \label{first-unique-label}
\end{figure} \\
Besides of the IVS correlations, the plantform also serves for the astrophysical and astrometric programs 
conducted with the Chinese VLBI network (CVN) and international joint VLBI observations. Meanwhile, the new-built
Tianma 65-m telescope will cover the frequency range from the L to the Q band together with two dual-frequency
receivers in S/X and X/Ka. The DiFX correlator successfully worked with the Tianma 65-m and found the fringes 
high to the X/Ka and Q  bands in late 2015.
\section{Performances and operations}
\subsection{Platform performances}
The computer cluster shown in Fig.1 is divided into two groups for the routine operations. Each head node manages
10 computing nodes, 200 CPU cores in total. The main features including the hardwares, the softwares and the 
network conditions are listed as follows. The maximum correlation speed is around 1 Gbps per station when processing
10 stations simultaneously (Figure 2). There are more than 6 staff (about 50\% working time) for different parts of 
the operations from data delivery to giving out the final outputs.
\begin{itemize}
 \item Correlator: DiFX-2.2/2.3/2.4/trunk \\
  Post-processing software: HOPS 3.9/3.10/3.11/3.12
 \item Head nodes: DELL R820 (E5-4610 CPU,2.4~GHz, 2*6 cores),64~GB Memory
                   DELL R730 (E5-2623 CPU,3.0~GHz, 2*4 cores),64~GB Memory
 \item Computing nodes: 20 DELL R630 nodes, 400 cores in total, 
                   2 socket Intel E5-2660 CPU(2.6~GHz, 10 cores),64~GB Memory
 \item I/O nodes: RAID6, 432~TB raw storage capacity
 \item Mark5 units: 3 Mark5A and 3 Mark5B.
 \item 56~Gb Infiniband for  internal computing network connection
 \item 1/10~Gb Ethernet for internal \& external network connection 
\end{itemize}

\begin{figure}[htb!]
  \includegraphics[width=.5\textwidth]{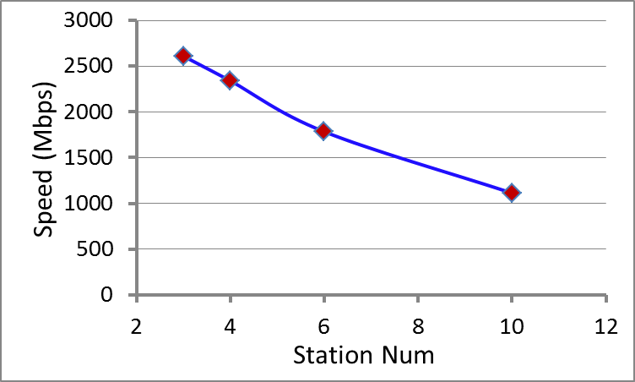}
  \caption{Correlation speed of Shanghai VLBI correlator.}
  \label{first-unique-label}
\end{figure}   
\subsection{e-transfer}
In order to process global IVS sessions, the network links to Fortzela,  HartRAO, Hobart, Kashima,
Noto, Sejong stations and Bonn correlator have been established (Figure 3). However, the links are not connected
in a real time mode, some time slots of conections are negotiated before the data transfer. The two Shanghai VLBI 
stations are in a 10 Gb link to the VLBI center while other CVN stations are in a much lower rate connection. Most of the 
high data rate and long duration recording experiments are still through shipment of the diskpacks in CVN. 
\begin{figure}[htb!]
  \includegraphics[width=.5\textwidth]{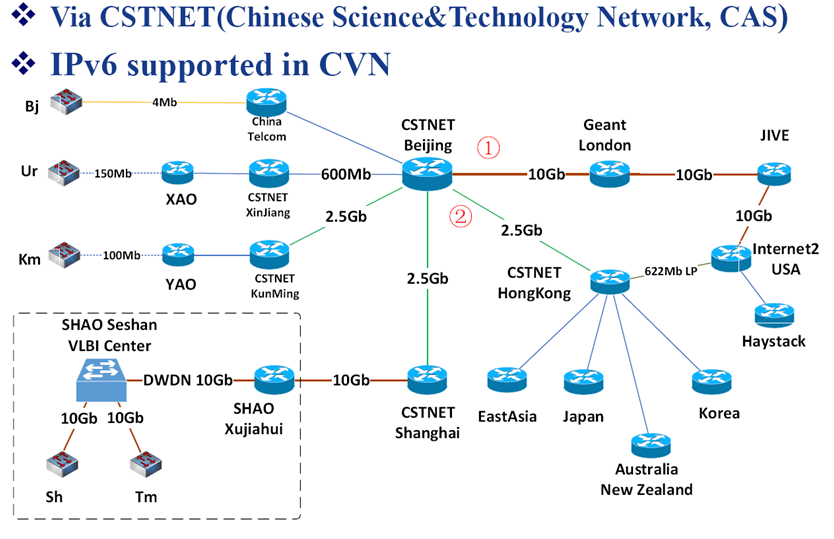}
  \caption{Network conditions at Shanghai VLBI center.}
  \label{first-unique-label}
\end{figure} 
\subsection{Statistics of correlation operations}
Some comparisons of the outputs after correlation and post-processing were made between the Shanghai DiFX correlator
and the Bonn DiFX correlator in late 2014. The RMS of the group delay differences in the X band extracted from 
an S/X session were within a few picoseconds. In early 2016, a similar 
comparison was made between the two correlators. The results listed in Table 1 implied the group delay of the two 
correlator outputs coincided at picosecond level.\\
Besides of serving for the global IVS sessions, the DiFX correlator is open to make correlations for the astrophysical and astrometric programs with CVN, east Asian and Australian joint VLBI observations. Table 2 
lists the summary of the correlations in details.\\
\begin{table*}[htb!]
\caption{Comparison results of Shanghai DiFX and Bonn DiFX correlators.}
\begin{tabular}{|l|c|c|c|c|c|r|} \hline
  & \multicolumn{3} {|c|} {S band} & \multicolumn{3} {|c|}{ X band }\\
\hline
Baseline & SNR \footnotemark[1]  & Group delay\footnotemark[2]  (ps) & Rate \footnotemark[2] (ps/s)  & SNR \footnotemark[1] & Group delay \footnotemark[2] (ps) &Rate\footnotemark[2]  (ps/s) \\
\hline
Ny-Ts & 1 & 2.4 & 0.0127 &   0.992 &1.1 &  0.0085     \\
Ny-Wn & 1 & 5.6 & 0.0208 &   1.002 &1.6 &  0.0095     \\
Ny-Wz & 1 & 3.8 & 0.0156 &   0.994 &0.9 &  0.0041     \\
Ts-Wn & 1 & 4.6 & 0.0198 &   1 &1.7 &  0.0091     \\
Ts-Wz & 1 & 2.8 & 0.0113 &   0.994 &0.9 &  0.0063     \\
\hline
\end{tabular}
\label{third-unique-label}
\end{table*}
\begin{table*}[htb!]
\caption{Summary of correlations processed.}
\begin{tabular}{|l|c|c|c|r|} \hline
Session Code & Observation Type  & Times in a year & Stations participated & Recording rate\\
\hline
 AUS-(AST,GEO) &   Geodesy       &12 (2016)  &Australian, more than 4 st.  &1024 Mbps  \\
 CVN-(CN) & Geodesy & 4 & CVN, Js and Ks\footnotemark[3], more than 3 st. & 512 Mbps   \\
 CVN-(PSR) & Pulsar Astrometry & not fixed &   CVN, 3 or 4 st.       & 1024 Mbps   \\
 CVN/EAVN & Astrophycis & not fixed & CVN or EAVN, more than 4 st. & 1024 Mbps  \\
 IVS-(AOV,APSG,CRDS,CRF,RD)  & Geodesy &\textgreater 10 & Global, up to more than 10 st. & 256/1024 Mbps  \\
 VEPS & Astrometry &  6 & east Asian and Australian, 3 or 4 st. & 2048 Mbps   \\
\hline
\end{tabular}
\end{table*}
\section{Some results}
\subsection{IVS and astrometric programs}
There were fifteen IVS sessions including eight CRF, three AOV, two APSG, one AUG and one CRDS series processed
and given out databases to the analysis center by the Shanghai correlator until now. The main time consumption was  
in the raw data delivery. 
Three CVN stations including Kunming 40-m, Shanghai 25-m and Urumqi 26-m participate ordinary IVS sessions.
The accuracy of their station positions achieves a few centimeters due to these long term global geodetic
sessions. It helps to carry out some astrometric programs based on the three stations. As also presented in 
this proceeding, an ecliptic plane survey program was based on the above three stations togther with one 
more stations from Hobart, Kashima and Sejong. The feasibilty of DiFX correlator made it possible to have
different quantifications and baseband bandwidth among different stations. In the first phase of observations, 
there were 435 target sources detected in three or more observations among more than 2000 candidates in the
 source pool. The detection rate was near 20\%.\\
A pulsar astrometry program has conducted with the S band receivers in the three stations. Five epoch 
phase-referenced VLBI positioning of the millisecond pulsar B1937+21 were carried out from 2012 to 2015.
The signal to noise ratio of the pulsar signal was improved by pulsar gating during the correlations. 
After EOP, station positions and ionospheric delay corrections, the best fitted proper motion in RA 
and DEC were 0.1237$\pm0.18$~mas$\slash$yr and -0.2585$\pm0.52$~mas$\slash$yr with a problematic 
parallax $\Pi$~=~-0.678~mas. Regardless of the parallax, the proper motion parameters were consitent 
with the 15.5 year timing solutions, 0.087(16)~mas$\slash$yr in RA and -0.41(3)~mas$\slash$yr in DEC. 
A deeper analysis is needed for the error mitigation.
\footnotetext[1]{The ratio of SNR .}
\footnotetext[2]{The WRMS of the differences in group delay and rate.}
\footnotetext[3]{Js: Jiamusi, Ks: Kashi, two deep space stations of China.}
\subsection{Tianma 65-m related}
The new-built Tianma 65-m radio telescope is about 6.1~km away from the Shanghai 25-m telescope. The receivers
installed make it have a continuous frequency coverage from the L band to the Q band. Two dual-frequency 
receivers in S/X and X/Ka bands play an important role in the geodetic activities. Besides of single dish
observations, Tianma 65-m is also an important site for the VLBI community. Some joint observations with
KaVA, EVN and VLBA have already been carried out in the low frequency bands. 
\begin{figure}[htb!]
  \includegraphics[width=.5\textwidth]{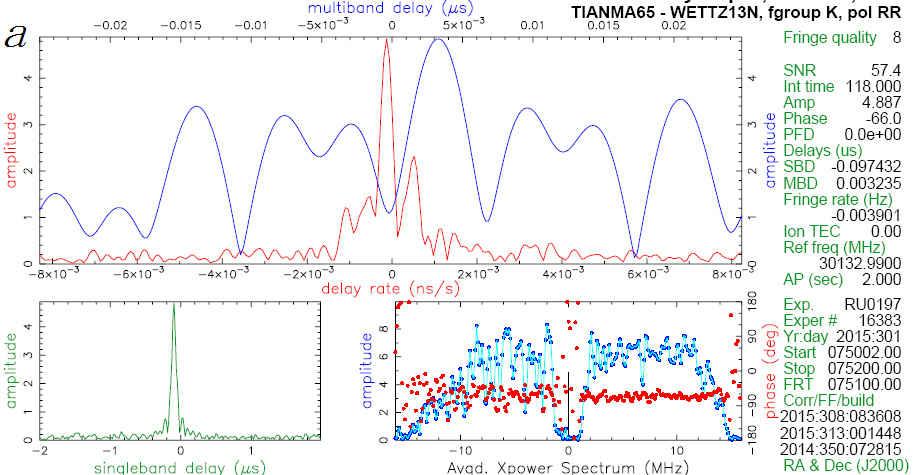} 
  \\
  \\
  \includegraphics[width=.5\textwidth]{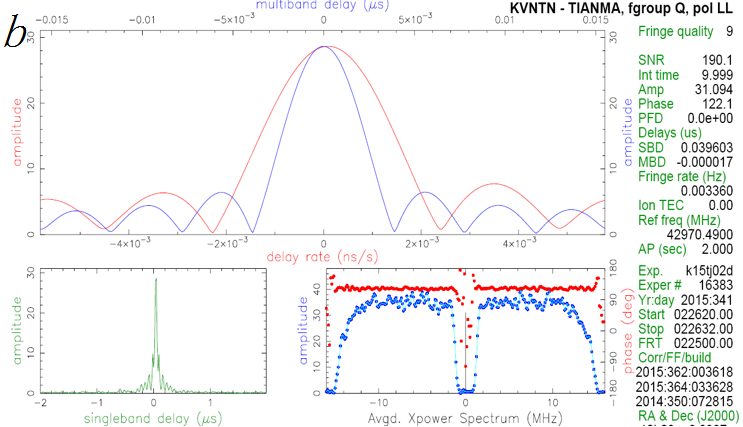}
  \caption{a. Ka (X$\slash$Ka) band fringes to Tianma 65-m. b. Q band fringes to Tianma 65-m.}
  \label{first-unique-label}
\end{figure} 
The fringes at high frequency bands including the X/Ka and the Q were found in late 2015 (Figure 5-a, 5-b).
The X/Ka experiment was carried out with Tianma-Wettz13n-Zelen13m in RU0197 session. While the Q band 
experiment was a Tianma-KaVA joint observation, an ad-hoc room temperature receiver was used in the experiment 
and the cooled dual-beam receiver is under installation in 2016. 
\section{Conclusions}
The DiFX plantform at SHAO is dedicated to the astrophysical and the geodetic VLBI observations. It serves as an
IVS correlator since 2015. The planform is also open to the CVN and joint international VLBI observations.
Concerning to the next generation broad band and dual polarization VLBI observations, the Shanghai
correlator will continue to make its contributions to the data correlation and processing.  
For the future high data rate and massive data correlations, current network condition will be one of the 
bottlenecks and must be improved. A more powerful platform with a high performance computing cluster and 
a competent storage system is also needed.
%

\section*{Acknowledgements}

The work is supported by the National Natural Science Foundation of China (No.Y247021001) and the 
Joint Funds of the National Natural Science Foundation of China (No.U1331205).

\begin{thebibliography}{99}
\bibitem{IVS-CC}
Deller A T, et al. PASP, 2011, 123(901): 275.
\bibitem{IVS-CC}
http://www.haystack.mit.edu/tech/vlbi/hops.html.
\bibitem{IVS-CC}
Reard on, D J, et al. MNRAS, 2016, 455(2):1751. 
\bibitem{IVS-CC}
Shu, F C, et al. \emph{IVS 2014 Annual Report}, 2014.
\end{thebibliography}
\end{document}